\documentclass{emulateapj}

\usepackage{graphicx}
\usepackage{natbib}

\shorttitle{Optical Classification of GRBs in the Swift Era}
\shortauthors{van der Horst et al.}

\begin{document}

\title{Optical Classification of Gamma-Ray Bursts in the {\it Swift} Era}

\author{ A.~J.~van~der~Horst\altaffilmark{1},
C.~Kouveliotou\altaffilmark{2}, 
N.~Gehrels\altaffilmark{3},
E.~Rol\altaffilmark{4}, 
R.~A.~M.~J.~Wijers\altaffilmark{5},
J.~K.~Cannizzo\altaffilmark{3,6}, 
J.~Racusin\altaffilmark{7},
D.~N.~Burrows\altaffilmark{7}, 
}

\altaffiltext{1}{NASA Postdoctoral Program Fellow, NSSTC, 
	320 Sparkman Drive, Huntsville, AL 35805; 
	Alexander.J.VanDerHorst@nasa.gov}
\altaffiltext{2}{NASA Marshall Space Flight Center, NSSTC, 
	320 Sparkman Drive, Huntsville, AL 38505}
\altaffiltext{3}{NASA Goddard Space Flight Center, 
	Greenbelt, MD 20771}
\altaffiltext{4}{Department of Physics \& Astronomy, University of Leicester, 
	University Road, Leicester, LE1 7RH, United Kingdom}
\altaffiltext{5}{Astronomical Institute, University of Amsterdam, 
	Kruislaan 403, 1098 SJ Amsterdam, The Netherlands}
\altaffiltext{6}{CRESST/Joint Center for Astrophysics, University of Maryland, 
	Baltimore County, Baltimore, MD, 21250}
\altaffiltext{7}{Department of Astronomy \& Astrophysics, Pennsylvania State University, 
	State College, PA 16802}

\begin{abstract} 
We propose a new method for the classification of optically dark gamma-ray bursts (GRBs), based on the
X-ray and optical-to-X-ray spectral indices of GRB afterglows, and
utilizing the spectral capabilities of {\it Swift}. This method depends
less on model assumptions than previous methods, and can be used as a quick
diagnostic tool to identify optically sub-luminous bursts. 
With this method we can also find GRBs that are extremely bright at optical wavelengths. 
We show that the previously suggested correlation between the optical darkness and the
X-ray/gamma-ray brightness is merely an observational selection effect.
\end{abstract}


\section{Introduction} \label{section:intro}

The discovery of gamma-ray burst (GRB) afterglows at optical wavelengths
\citep{vanparadijs1997:nature} led to the confirmation of the
extragalactic nature of the phenomenon \citep{metzger1997:nature}. In
the last decade, a multitude of photometric and spectroscopic
observations have established the relativistic blast wave model for the
afterglow emission \citep{rees1992:mnras, wijers1997:mnras}. 
Coordinated multi-wavelength observations have
revealed the nature of a large fraction of GRB progenitors
\citep{galama1998:nature, hjorth2003:nature, stanek2003:apj} and
allowed studies of their host galaxies \citep{sahu1997:nature}. In very
few cases, however, GRB counterpart searches were not successful in
identifying a fading transient at the GRB explosion site. Only a few months after
the first afterglow detection, \cite{groot1998:apj} found
the first optically dark event, GRB\,970828. They used the concept of 'dark
burst' to indicate that the event was detected in X-rays but
not in the optical, down to deep limiting magnitudes at a few hours
after the burst. Since then, observers classified many bursts as dark,
although in a substantial fraction of these the absence of an optical
counterpart could simply be due to a lack of observational sensitivity
or a large delay between the GRB trigger time and the start of the
follow up observations. Even excluding these cases, there remain a few
events that do not fall within these observational constraints. Several
explanations have been proposed in the literature for their dark nature:
they could be intrinsically faint at optical wavelengths compared to the
X-rays, or they might reside at high redshifts, or they might be obscured
by gas and dust in their host galaxies \citep[e.g.][]{rol2005:apj}.

To get a better handle on the optical darkness of GRB afterglows by
eliminating some of the observational effects, two major efforts were
undertaken before the launch of NASA's {\it Swift} GRB satellite. Both
constructed a consistent method to define what classifies a GRB as being
dark. \cite{jakobsson2004:apj} defined dark bursts solely based on their
optical-to-X-ray emission spectral index being lower than a certain
theoretical value of 0.5. \cite{rol2005:apj} applied a more elaborate
approach, extrapolating the observed X-ray flux to the optical regime by
using their X-ray spectral and temporal indices and assuming that the GRB 
explosion followed the relativistic blast wave model. Both
approaches were demonstrated on samples of pre-{\it Swift} GRBs and
resulted in largely overlapping dark GRB sets.

In this paper we examine the identification of dark bursts in the light
of the new observational window in the X-rays and the optical 
wavelengths opened by the {\it Swift} satellite, seconds after the
detection of the GRB prompt emission. We have studied the sample of long 
{\it Swift} GRBs from \cite{gehrels2008:apj} with X-ray and optical
detections or stringent upper limits (Tables 2 and 3 from
\cite{gehrels2008:apj}; optical fluxes from ground-based telescopes 
and Swift's UV Optical Telescope; X-ray spectral indices from \cite{racusin2008:arxiv}). 
We added to the sample two {\it Swift} GRBs:
GRB\,050401 \citep{depasquale2006:mnras, watson2006:apj} and GRB\,050410
\citep{mineo2007:aa}, with an optical detection and an optical upper
limit respectively. We also added two {\it HETE-II} GRBs: GRB\,051022
\citep{nakagawa2006:pasj, rol2007:apj, castrotirado2007:aa} and
GRB\,051028 \citep{castrotirado2006:aa, urata2007:pasj}, with an optical
upper limit and an optical detection respectively. To conform with the
\cite{gehrels2008:apj} data set we inter- or extrapolated the X-ray and
optical fluxes of these last four GRBs to 11 hours and corrected for
galactic extinction.

In Section \ref{sec:current} we apply and discuss the current dark burst
classification schemes to this entire sample. We propose a new method in
Section \ref{sec:newmethod}, which utilizes the unique capabilities of
{\it Swift}. In Section \ref{sec:individual} we briefly discuss the GRBs
that are classified as optically dark by our method. Section
\ref{sec:obseffect} examines the apparent correlation between optical
darkness and X-ray/gamma-ray brightness; and we summarize our results in
Section \ref{sec:conclusions}.

\section{Current dark burst classification methods}\label{sec:current}

The {\it Swift} GRB afterglow data are unprecedented in that they start typically within two minutes after each GRB trigger and monitor temporal evolution with high resolution and for long periods (from days to tens of days). Several afterglow studies have now established that GRB X-ray light curves are much more complex than previously thought: the canonical {\it Swift} X-ray light curve displays additional (earlier) phases of steep and
slow decay on what was earlier considered the 'normal' decay \citep{nousek2006:apj, zhang2006:apj}. Various explanations have been put forward for this behavior \citep[e.g][]{granot2006:mnras, panaitescu2006:mnras}; at the same time it is becoming obvious that simply extrapolating the
X-ray flux to the optical bands using {\it temporal} decay indices and predictions
from the standard blast wave model is not straightforward and should be done with great caution.
Extrapolating the X-ray flux to the optical bands by adopting their X-ray
{\it spectral} index, however, is more sensible since it only
assumes that the broadband spectral energy distribution is caused
by synchrotron radiation, which is well established for GRB afterglows. We note here that
Inverse Compton radiation might also have a small effect in the X-ray spectrum, 
as it could make the burst optically sub-luminous compared to the X-ray emission. 

According to the above, both dark-GRB determination methods, by \cite{jakobsson2004:apj} and \cite{rol2005:apj}, have their drawbacks, and we believe that an improvement can now be made
with the new results that {\it Swift} has provided. The method of \cite{rol2005:apj} is most affected, as it uses both spectral and temporal indices, while the \cite{jakobsson2004:apj} method requires no temporal evolution assumptions. 
In the latter method, the optical and X-ray fluxes are interpolated or extrapolated to 11 hours after the burst and plotted together with a line of constant optical-to-X-ray spectral index $\beta_{\rm{OX}}$; bursts that fall below that line are classified as dark.  
The drawback of this method, which has also been pointed out in \cite{jakobsson2004:apj}, is the choice of the dividing $\beta_{\rm{OX}}$; they chose a value of 0.5 as a minimum based on the following arguments. In synchrotron radiation, the photon spectral index in the optical and X-rays depends on the power-law index $p$ of the electron energy distribution as either $(p-1)/2$ or $p/2$, if the cooling frequency is situated above or below the observing band, respectively. In the simplest form of the blast wave model \citep[e.g.][]{wijers1997:mnras, sari1998:apj}, $p$ is expected to be larger than $2$ leading to $\beta_{\rm{OX}} \geq 0.5$. In most cases, $p$ is indeed found to be $>2$ leading to $\beta_{\rm{OX}} \geq 0.5$, but it has been shown that $p<2$ is possible, both observationally and theoretically, by introducing a high-energy cutoff in the electron energy distribution \citep{dai2001:apj, panaitescu2002:apj, bhattacharya2004:aspcs, starling2008:apj}.

\section{Our classification method}\label{sec:newmethod}

\begin{figure*}
\begin{center}
\includegraphics[angle=-90,width=\textwidth]{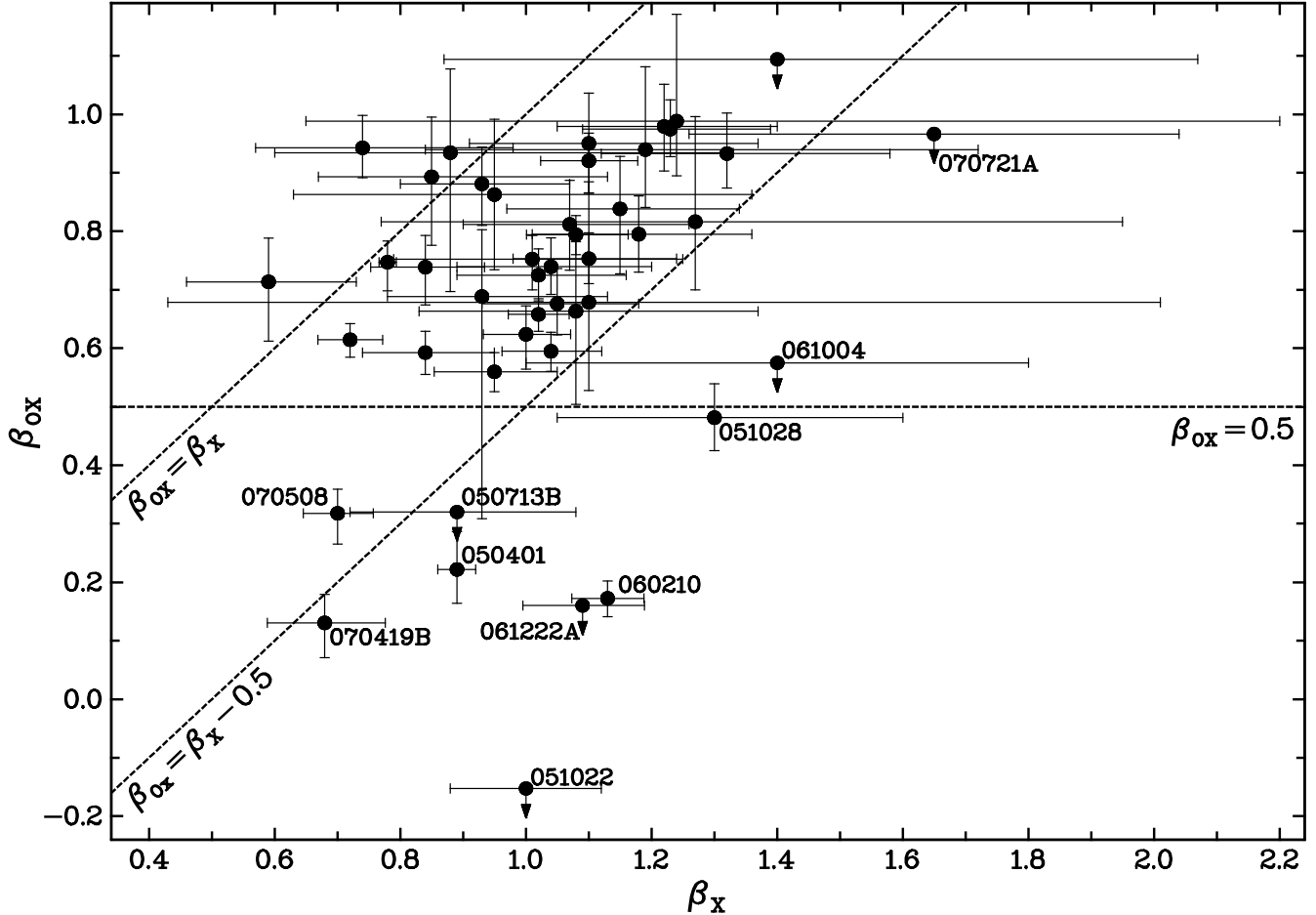} \caption{The
spectral index $\beta_{\rm{OX}}$ between the optical and X-ray observing
bands versus the X-ray spectral index $\beta_{\rm{X}}$ for our sample of
{\it Swift} GRBs (error bars at 90\% and upper limits at $3\sigma$ level). 
The optical-to-X-ray spectral index is given at 11
hours after the burst. The horizontal dashed line indicates the dark
burst criterion from \cite{jakobsson2004:apj}: the optical-to-X-ray
spectral index $\beta_{\rm{OX}}$ is equal to $0.5$. The diagonal dashed
lines indicate $\beta_{\rm{OX}}=\beta_{\rm{X}}$ and
$\beta_{\rm{OX}}=\beta_{\rm{X}}-0.5$.} 
\label{fig:betaoxvsbetax}
\end{center}
\end{figure*}

We propose a new method for dark burst classification which does not involve the temporal index of the X-ray afterglow light curves and has fewer assumptions regarding the emitting electron energy distribution. Our main assumptions are that the underlying emission mechanism in both the optical and X-rays is synchrotron radiation, and that the emission in both frequency bands originates from the same source. In that case, given a certain value for the X-ray spectral index $\beta_{\rm{X}}$, the optical spectral index $\beta_{\rm{O}}$ should either have the same value or a value of $\beta_{\rm{X}}-0.5$, if there is a cooling break in between the optical and X-ray regimes. This implies that $\beta_{\rm{OX}}$, the spectral index between the X-ray and optical regime, should be in between $\beta_{\rm{X}}$ and $\beta_{\rm{X}}-0.5$. In Figure \ref{fig:betaoxvsbetax} we plot $\beta_{\rm{OX}}$ versus the X-ray spectral index $\beta_{\rm{X}}$ for the {\it Swift} GRB sample from \cite{gehrels2008:apj} (uncertainties in parameters at 90\% and upper limits at $3\sigma$ confidence level). In this method, again $\beta_{\rm{OX}}$ is constructed from the interpolated/extrapolated optical and X-ray flux at 11 hours; moreover, $\beta_{\rm{X}}$ is quickly available with good accuracy from the {\it Swift} XRT for almost all GRBs. It is necessary to take these spectral indices at late times, i.e. after several hours, 
since some GRBs have bright optical emission at early times, presumably from reverse shocks.

In principle, all GRBs should be found in a band between the lines of $\beta_{\rm{OX}} = \beta_{\rm{X}}$ and $\beta_{\rm{OX}}=\beta_{\rm{X}}-0.5$. There are two regions of interest in Figure \ref{fig:betaoxvsbetax}:  GRBs with brighter than expected optical emission at the left upper part of the Figure, 
and GRBs where $\beta_{\rm{OX}}$ is even shallower than $\beta_{\rm{X}}-0.5$. 
We define these latter bursts to be optically sub-luminous, or dark, in our classification method. 
The bursts for which the error bars on $\beta_{\rm{X}}$ overlap with the line 
$\beta_{\rm{OX}}=\beta_{\rm{X}}-0.5$ are not classified as dark in our method.
The resulting dark burst population in our sample is different than the one obtained applying the criterion of \cite{jakobsson2004:apj}, which in Figure \ref{fig:betaoxvsbetax} are those GRBs below the horizontal line $\beta_{\rm{OX}}=0.5$. Also, we note that in our sample there are three events where there is clearly a detection in the optical wavelengths, defying their classification of dark as such. We discuss the individual dark bursts as well as the optical bright events in the next section. 

Further, one of the main differences between \cite{jakobsson2004:apj} and our method is that we accommodate here all values that are found for the various spectral indices, and thus all implied values of $p$. Thus, bursts with a shallow X-ray spectrum (indicating a low value for $p$), can still be `normal', i.e. not dark, in our method, while they could be classified as being dark in the method of \cite{jakobsson2004:apj}. Indeed, as was already pointed out by \cite{jakobsson2004:apj}, the fact that a certain burst has a low value of $p$ should not have any implications for its `darkness' classification, especially given that various observational studies have shown that there is  not a single universal value of $p$ but a wide distribution of values \citep[e.g.][]{panaitescu2002:apj, starling2008:apj}. Conversely, if a GRB has a steep X-ray spectrum, the value for $\beta_{\rm{OX}}$ should also be high, and if this is not the case it could be classified as a dark burst. The latter is not accommodated in the \cite{jakobsson2004:apj} method; it is correctly predicted in the method by \cite{rol2005:apj}, although there again the temporal indices are the unreliable factor in the classification.

\section{Comments on individual bursts}\label{sec:individual}

Figure \ref{fig:betaoxvsbetax} shows that there are several GRBs outside of the band between the lines of $\beta_{\rm{OX}} = \beta_{\rm{X}}$ and $\beta_{\rm{OX}}=\beta_{\rm{X}}-0.5$, which we will discuss here in some detail. 

\subsection{Optically dark bursts}
Looking at Figure \ref{fig:betaoxvsbetax}, we notice that there are some GRBs which are clearly classified as dark by the method of \cite{jakobsson2004:apj} and by our method, but there are also some differences. GRBs 051022, 061222A, 050401, and 060210 are classified as dark by both methods; 
the first two are optical upper limits, and the
latter two are actually detections but optically sub-luminous events. 
GRBs 050713B, 070508, and 070419B are classified as dark by \cite{jakobsson2004:apj} 
since they have $\beta_{\rm{OX}}<0.5$, but not according to our method; 
given the low value for $\beta_{\rm{X}}$, these are cases of $p<2$. 
GRB\,051028 is identified as dark by our method, although the uncertainty in $\beta_{\rm{X}}$ is rather large, and is not identified as such by the \cite{jakobsson2004:apj} method. 
Finally, GRBs 061004 and 070721A have optical upper limits and are not classified as dark by the \cite{jakobsson2004:apj} method, and due to their large uncertainties in $\beta_{\rm{X}}$ also not by our method. 
We note that the latter two are situated in the region of the diagram of possible dark bursts with a high value of $p$; given their large errors in the determination of $\beta_{\rm{X}}$, we will not discuss these particular two events further.

Detailed analyses of GRB\,050401 \citep{depasquale2006:mnras,
watson2006:apj} and GRB\,060210 \citep{curran2007:aa} show that the sub-luminous optical flux of these events can be explained by a combination of moderate local (host galaxy) extinction and high redshift (2.90 and 3.91, respectively).
GRB\,051022 was detected at X-ray, millimeter and radio frequencies, and
optical observations showed a moderately bright host galaxy but no
optical afterglow emission down to deep limits \citep{rol2007:apj,
castrotirado2007:aa}. Broadband modeling of the afterglow by
\cite{rol2007:apj} constrained not only the physical parameters of the
blast wave, but also the severe host galaxy extinction of this
relatively nearby ($z=0.809$) GRB. 
The afterglows of GRBs 051028 and 070419B were detected in the optical, but there are no spectroscopic redshifts
available; and also not for GRB 050713B and 061222A, which are not detected in the optical. There is not enough data available to investigate these events further, although we note that for GRB\,051028 a relatively high redshift is the most probable cause for it being (marginally) optically sub-luminous \citep{castrotirado2006:aa}. 

\subsection{Optically bright bursts}
The left upper part of Figure \ref{fig:betaoxvsbetax}, above the $\beta_{\rm{OX}} = \beta_{\rm{X}}$ line, 
is of interest, because it reveals GRBs that are optically super-luminous. 
In our sample two events have $\beta_{\rm{OX}}$ values above the $\beta_{\rm{OX}} = \beta_{\rm{X}}$ line, 
GRB\,060607A and GRB\,060526, but they are consistent with $\beta_{\rm{X}}$ 
at the 90\% confidence level. 
Two more are marginally above, but consistent with, that line, namely GRB\,050908 and GRB\,050801. 
All these four events are at the low end of the $\beta_{\rm{X}}$ distribution in our sample, with GRB\,060607A actually having the lowest $\beta_{\rm{X}}$ value of the whole sample. 
 
These optically bright sources are consistent with $\beta_{\rm{OX}} = \beta_{\rm{X}}$ 
within measurement uncertainties. 
However, three of these sources are at a high redshift ($z=3.08$, $3.22$, and $3.34$), 
and a moderate host galaxy extinction would increase the value of $\beta_{\rm{OX}}$ in Figure \ref{fig:betaoxvsbetax} while $\beta_{\rm{X}}$ remains at the same value, 
i.e. the source would move up vertically in this Figure away from the $\beta_{\rm{OX}} = \beta_{\rm{X}}$ line. 
This would make the brighter than expected optical brightness of these sources more significant. 
We note that a change in $\beta_{\rm{OX}}$ due to a host galaxy extinction of $A_{\rm{V}}$ magnitudes 
can be approximated by $0.13\cdot A_{\rm{V}}$. 
For an average extinction of $\sim 0.2$ magnitudes in the source frame, 
as found for a pre-{\it Swift} sample of GRB host galaxies \citep{kann2006:apj}, this effect is negligible. 
However, some studies \citep[e.g.][]{cenko2009:apj} show that several individual host galaxies display strong 
extinction with $A_{\rm{V}}\sim 1$ or larger, which does have a significant effect on the value of $\beta_{\rm{OX}}$.

With regards to the spectral indices of GRBs in that part of Figure \ref{fig:betaoxvsbetax}, 
there are various possibilities for their deviating values. 
Inverse Compton radiation could play a role at X-ray frequencies, which would flatten the X-ray spectral index but also increase the X-ray flux and thus decrease $\beta_{\rm{OX}}$. 
It is not obvious that correcting for this effect would move these sources into the allowed band of spectral indices in Figure \ref{fig:betaoxvsbetax}. 
A second possible explanation for the extreme optical brightness could be that the spectrum is not described by one broadband spectrum, but that there is an extra emission component at optical wavelengths. 
This could be caused by a double-jet configuration or refreshed shocks, which have both been invoked for explaining some well-sampled GRB afterglows \citep[e.g.][]{berger2003:nature,starling2005:mnras}.

\section{Optical darkness versus X-ray and gamma-ray brightness}\label{sec:obseffect}

\begin{figure}
\begin{center}
\includegraphics[angle=-90,width=0.75\columnwidth]{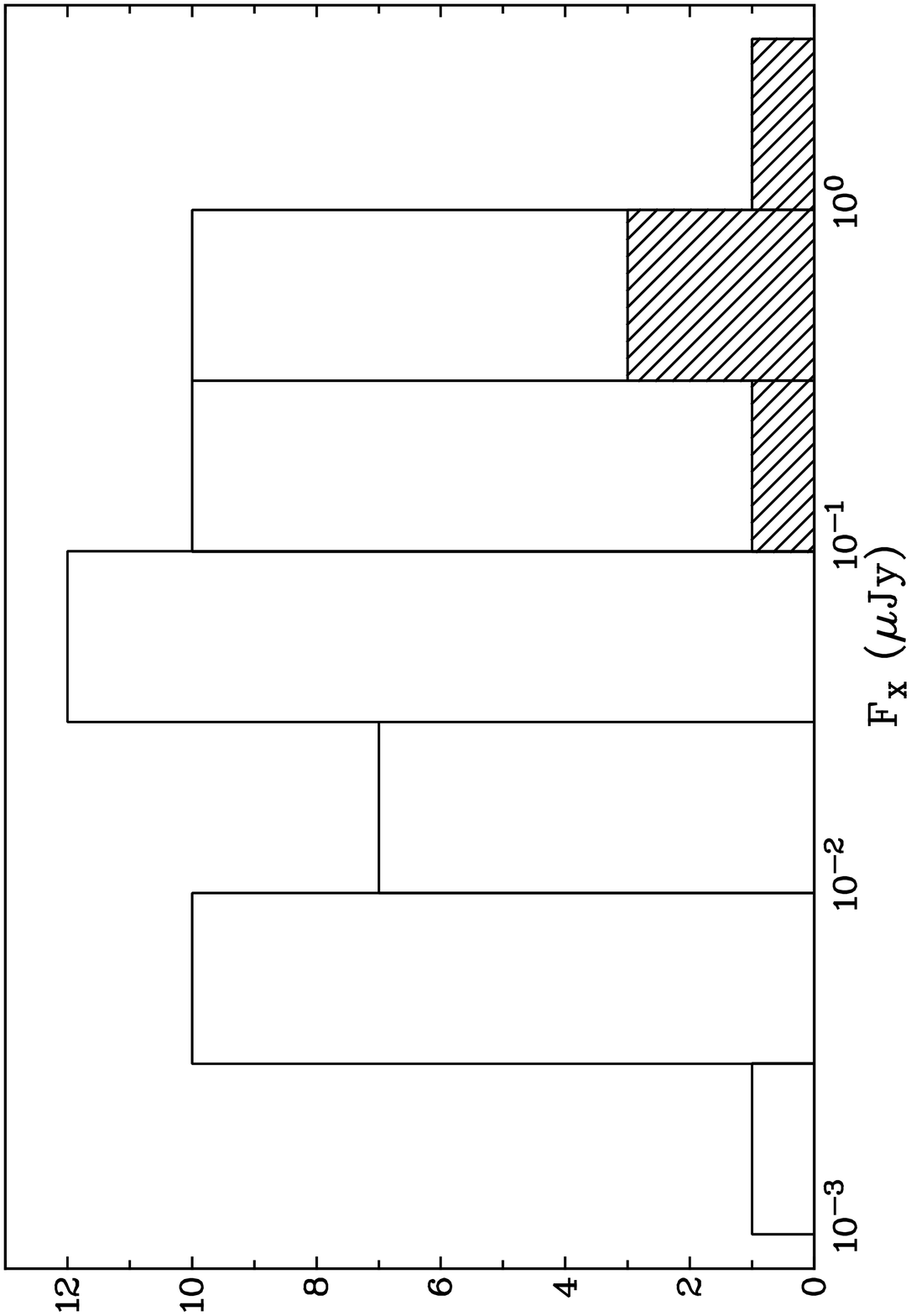}
\includegraphics[angle=-90,width=0.75\columnwidth]{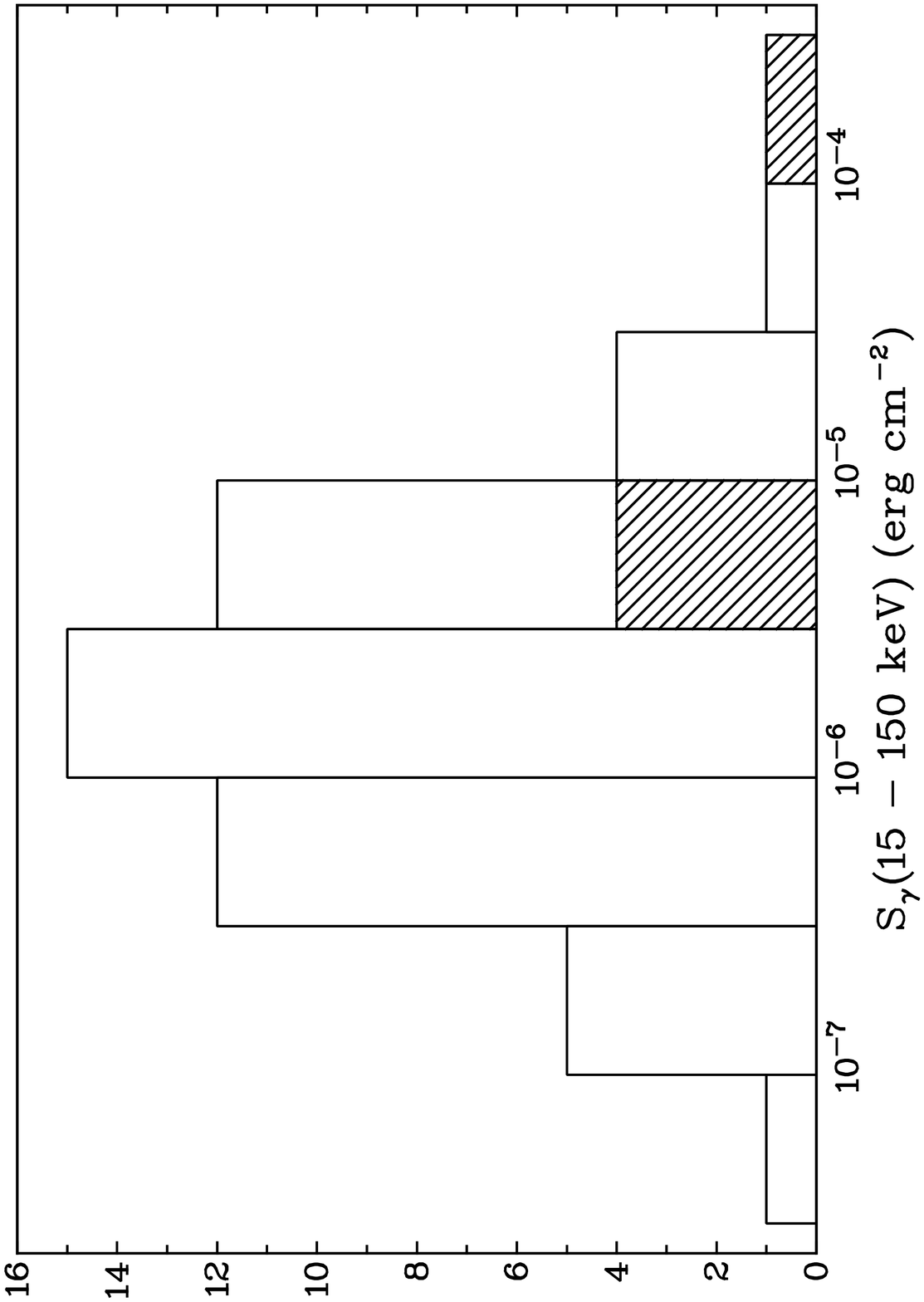} \caption{Histogram
of the X-ray flux $F_{\rm{X}}$ (top) and the gamma-ray fluence
$S_{\gamma}$ in the 15-150 keV energy range (bottom) for our total sample
of  {\it Swift} GRBs. The GRBs classified as dark bursts by our method
are indicated by double hatched histograms. These
histograms suggest that almost all optically dark GRBs are bright in X-rays and relatively bright in gamma-rays.}
\label{fig:histof}
\end{center}
\end{figure}

We compare below the X-ray and gamma-ray brightness of the GRBs classified as optically sub-luminous with our method, to our whole sample. Figure \ref{fig:histof} shows the distributions of their X-ray flux $F_{\rm{X}}$, and gamma-ray fluence $S_{\gamma}$ in the $15-150$~keV range. 
The gamma-ray fluences of the two {\it HETE-II} bursts, GRB\,051022 and GRB\,051028, have been converted to values in the correct energy range by using the fluences and spectral parameters provided in \cite{nakagawa2006:pasj} and \cite{golenetskii2005:gcn}, respectively. 
Figure \ref{fig:histof} strongly suggests that optically dark GRBs are bright in X-rays and relatively bright in gamma-rays, as also suggested earlier by \cite{depasquale2003:apj} and \cite{rol2005:apj}. 
We have explored the possibility of an observational selection effect to explain this latter correlation: 
if there is a positive correlation between optical and X-ray/gamma-ray brightness, 
it is easier to identify optically dark bursts if they are bright in X-rays and gamma-rays. 
Here we show that this selection effect in indeed in play.

\begin{figure*}
\begin{center}
\includegraphics[angle=-90,width=\textwidth]{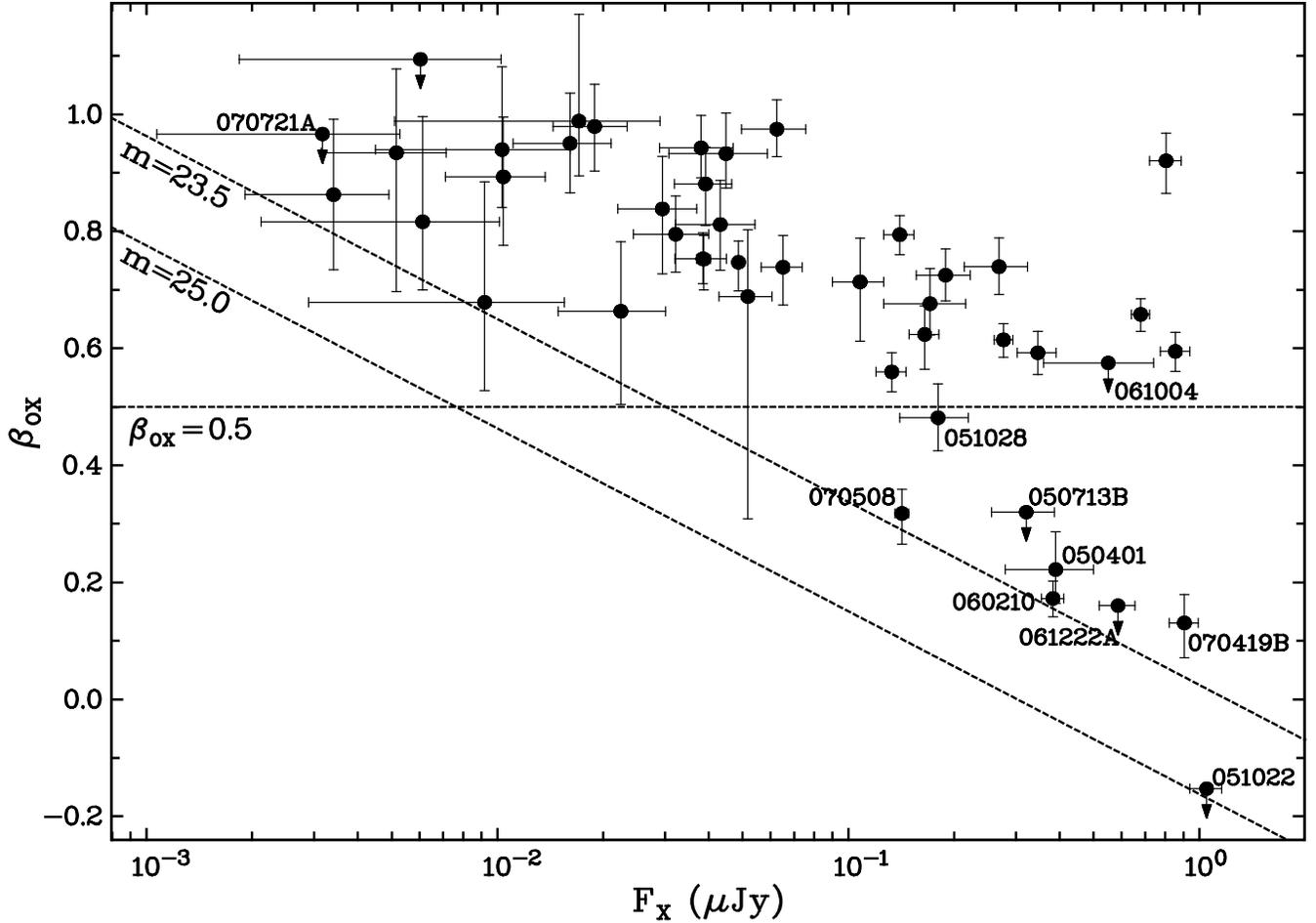} \caption{The
spectral index $\beta_{\rm{OX}}$ between the optical and X-ray observing
bands versus the X-ray flux $F_{\rm{X}}$ for our sample of {\it
Swift} GRBs (error bars at 90\% and upper limits at $3\sigma$ level). 
Both the spectral index and the X-ray flux are given at 11
hours after the burst. The horizontal dashed line indicates the dark
burst criterion from \cite{jakobsson2004:apj}: the optical-to-X-ray
spectral index $\beta_{\rm{OX}}$ is equal to $0.5$. The diagonal dashed
lines indicates optical magnitudes of 23.5 and 25, showing that 
optically dark GRBs being bright in X-rays is merely an observational
selection effect.}
\label{fig:betaoxvsfx}
\end{center}
\end{figure*}

Figure \ref{fig:betaoxvsfx} exhibits $\beta_{\rm{OX}}$ versus X-ray flux; the dashed lines indicate
optical magnitudes of 23.5 and 25, which is roughly the limiting magnitude
range of the current deepest searches for optical afterglows. We see here the same trend as that seen in the top
histogram of Figure \ref{fig:histof}, namely the correlation between
optical darkness and X-ray brightness. Figure \ref{fig:betaoxvsfx} shows that there are no bursts below the magnitude 25
line, and only one between 23.5 and 25 (GRB\,051022). Above magnitude 23.5 the GRBs are quite homogeneously
distributed, clearly suggesting an observational selection effect. This
only explains the selection effect for the correlation between the
optical darkness and X-ray brightness. The correlation between the
optical darkness and gamma-ray fluence can be understood from a
different correlation, studied for this sample by
\cite{gehrels2008:apj}, namely the positive correlation between the
X-ray flux and gamma-ray fluence.

\section{Conclusions}\label{sec:conclusions}

We have proposed a new method for the classification of optically dark
gamma-ray bursts (GRBs), based on the X-ray and optical-to-X-ray
spectral indices of GRB afterglows, and utilizing the spectral
capabilities of {\it Swift}. When plotting the optical-to-X-ray spectral
index $\beta_{\rm{OX}}$ versus the X-ray spectral index
$\beta_{\rm{X}}$, all the GRBs below the dividing line of
$\beta_{\rm{OX}}=\beta_{\rm{X}}-0.5$ are classified as dark. This method
depends less on model assumptions than previous methods, and can be used
as a quick diagnostic tool to identify dark bursts. 
In particular, the method assumes only a synchrotron spectrum from optical to X-rays originating in one emission region, 
allows for the large range of observed values for the electron energy distribution index $p$, 
and excludes any uncertainties in the temporal evolution of the GRB blast wave. 
With our classification method we can also find GRBs that are extremely bright in the optical, 
i.e. in which $\beta_{\rm{OX}}$ is too steep compared to $\beta_{\rm{X}}$. 
We suggest that this could be due to other radiation processes besides synchrotron radiation at X-ray frequencies, 
e.g. Inverse Compton; or due to extra emission in the optical, for example caused by a double-jet configuration or refreshed shocks. 

Utilizing our sample of {\it Swift}, and two {\it HETE-II}, GRBs, we have shown that the previously suggested correlation 
between the optical darkness and the X-ray/gamma-ray brightness is merely an observational selection effect. 
In order to further the understanding of dark bursts, in particular the effect of host galaxy extinction on the blast wave emission, 
it is important to undertake dedicated searches for dark burst candidates from early-time optical, UV and (near-)infrared observations. 
Since the afterglow physics in the first minutes to hours after the burst is quite uncertain, 
these early-time observations should be followed up by observations at roughly half a day after the burst with 
at least 4-meter class telescopes, in order to constrain the optical darkness as stringent as possible. 
Constraining the full broadband spectrum by including radio and millimeter observations, 
as for example done for GRB\,051022, could then provide important clues on the nature of dark bursts and their host galaxies.

\acknowledgements{\small We thank Pall Jakobsson for useful discussions and suggestions. 
AJvdH was supported by an appointment to the NASA Postdoctoral Program at the MSFC, 
administered by Oak Ridge Associated Universities through a contract with NASA. 
JR and DNB gratefully acknowledge support from NASA contract NAS5-00136.}

\bibliographystyle{apj}
\bibliography{references}

\end{document}